\documentclass[journal, twoside, onecolumn, draftclsnofoot,12pt]{IEEEtran}

\usepackage{multirow}  \usepackage{cite}
\usepackage{graphicx,subfigure}\usepackage{amsmath} \usepackage{amssymb} \usepackage{amsthm}\usepackage{color,array}
\usepackage{etex,etoolbox} \usepackage{float}
\usepackage{color}\usepackage{graphicx}
\usepackage{caption}\usepackage{xcolor}
\usepackage{colortbl}
\usepackage{soul}
\usepackage{algorithm}
\usepackage{soul,color,bm}
\usepackage{setspace}

\usepackage{algorithmic}   \usepackage{bbm}
\usepackage[T1]{fontenc}

\makeatletter
\providecommand{\@fourthoffour}[4]{#4}
\def\fixstatement#1{%
  \AtEndEnvironment{#1}{%
    \xdef\pat@label{\expandafter\expandafter\expandafter
      \@fourthoffour\csname#1\endcsname\space\@currentlabel}}}

\globtoksblk\prooftoks{1000}
\newcounter{proofcount}
\stepcounter{proofcount}
\long\def\proofatend#1\endproofatend{%
  \edef\next{ \Alph{proofcount}. Proof of \pat@label \noexpand\begin{proof}[Proof]}%
  \toks\numexpr\prooftoks+\value{proofcount}\relax=\expandafter{\next#1\end{proof}}
  \stepcounter{proofcount}}

\def\printproofs{%
  \count@=\z@
  \loop
    \the\toks\numexpr\prooftoks+\count@\relax
    \ifnum\count@<\value{proofcount}%
    \advance\count@\@ne
  \repeat}

\makeatother

\fixstatement{thm}
\fixstatement{lem}

\begin{document}

%%%%%%%%%%%%%%%%%%%%%%%%%%%%%%%%%%%%%%%%%%%%%%%%%%%%%%%%%%%%%%%%%%%%%%%
% Title
%%%%%%%%%%%%%%%%%%%%%%%%%%%%%%%%%%%%%%%%%%%%%%%%%%%%%%%%%%%%%%%%%%%%%%%

\title{Application of End-to-End Deep Learning in Wireless Communications Systems}
\author{Woongsup Lee, Ohyun~Jo, and Minhoe Kim
\thanks{W. Lee is with the Department of Information and Communication Engineering, Gyeongsang 
National University, South Korea. 
O. Jo is with the Department of Computer Science, Chungbuk National University, South Korea.
M. Kim is with the Department of Communication 
Systems, EURECOM, 06410 Sophia-Antipolis, France. This work has been submitted to the IEEE for possible publication. Copyright may be transferred without notice, after which this version may no longer be accessible.}}

\maketitle

%\thispagestyle{empty}
%\pagestyle{empty}
%\thispagestyle{plain}\pagestyle{plain}
%\centerfigcaptionstrue

%\maketitle
%%%%%%%%%%%%%%%%%%%%%%%%%%%%%%%%%%%%%%%%%%%%%%%%%%%%%%%%%%%%%%%%%%%%%%%
% use to conference
%%%%%%%%%%%%%%%%%%%%%%%%%%%%%%%%%%%%%%%%%%%%%%%%%%%%%%%%%%%%%%%%%%%%%%%

\begin{abstract}

Deep learning is a potential paradigm changer for the design of wireless communications 
systems (WCS), from conventional handcrafted schemes
based on sophisticated mathematical models with assumptions to autonomous schemes based 
on the end-to-end deep learning using a large number of data. In this article, we present 
a basic concept of the deep learning and its application to WCS by investigating the
resource allocation (RA) scheme based on a deep neural network (DNN) where 
multiple goals with various constraints can be satisfied through the
end-to-end deep learning. Especially, the optimality and feasibility of the DNN based RA are
verified through simulation. Then, we discuss the technical challenges 
regarding the application of deep learning in WCS.

\end{abstract}

\section{Introduction} 

%%%%%% 딥러닝에 대한 이야기 시작
%%%%%% -> 딥러닝이 요즘 잘나감
%%%%%% -> 이미지넷과 같은 이미지 분류 및 스피치 인식에서 압도적인 성능
%%%%%% -> 그외에도 질병탐색, 등등의 다양한 분야에서 활용됨
%%%%%% -> 통신에서도 최근 활발하게 활용되고 있음

The deep learning, which is based on the deep neural 
network (DNN) that emulates the neurons of the brain, 
has gained in great popularity in recent days. The current enthusiasm 
for deep learning is mainly due to its significant 
performance gains over conventional schemes\cite{Shea2016, Karpathy2015, LeCun2015}.
For example, the deep learning based image classification
schemes can achieve far more accurate performance than 
handcrafted conventional schemes based on the analytic models,
and they have even surpassed the human-level performance. 
The application of deep learning is not confined to the simple classification 
task but it also shows notable performance in 
more complicated tasks, such as the semantic scene 
understanding \cite{Karpathy2015}.

\begin{figure}[t]
	\centerline{\includegraphics[width=16.0cm]{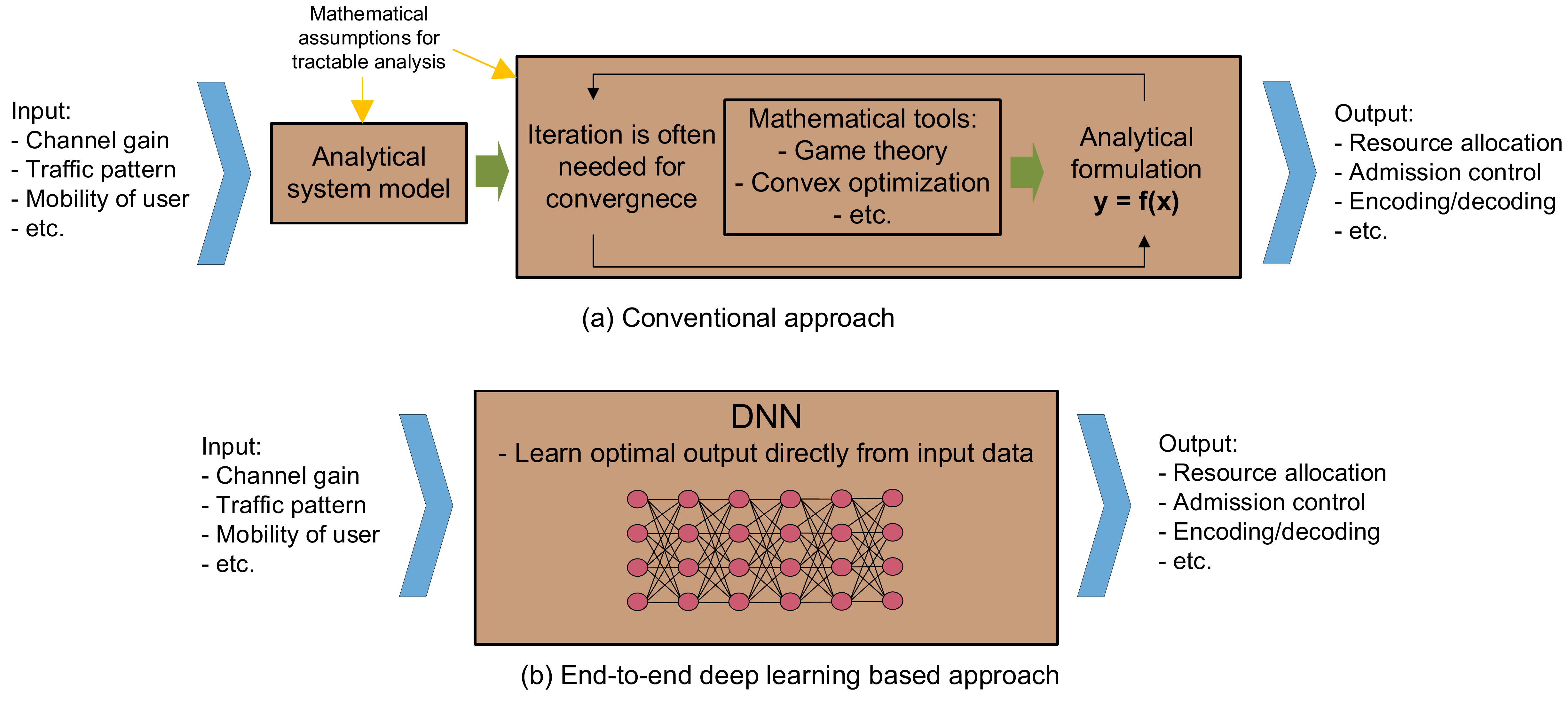}}
	\caption{Comparison of end-to-end deep learning based approach with conventional approach.}
	\label{fig_e-to-e-learn}
\end{figure}

The advent of deep learning can change the research paradigm from 
designing scheme through careful engineering based on mathematical models
to end-to-end learning based scheme in which the proper scheme
is autonomously designed by observing a large amount of data, cf. Fig. \ref{fig_e-to-e-learn}.
For example, conventional image classification schemes rely
on the handcrafted complex feature detectors which are engineered by the expertise, 
e.g., edge detector. However, in case of deep learning based 
image classification, feature detectors that are 
far more accurate than conventional detectors can be derived by 
a DNN structure from a large number of image data.
Accordingly, in the era of deep learning, 
1) the preparation, selection and pre-processing
of data to be used in DNN structure, 2) the determination of proper DNN 
structure and 3) the interpretation of the output of DNN, become more
important than the development of analytic schemes from
the mathematical system model which usually contains 
assumption to make analysis feasible.
%==============================================================
%==============================================================

In recent days, the deep learning has begun to be applied to the 
many research areas of wireless communications systems (WCS),
especially in the classification tasks such as traffic and channel estimation.
The authors of \cite{OShea2016a} showed that the
type of data traffic can be determined accurately with DNN.
Moreover, in \cite{Ye2018}, the channel estimation and signal 
detection in orthogonal frequency-division multiplexing (OFDM) systems was
considered based on the deep learning. Furthermore, the 
deep learning is also applied to more complicated tasks than just classification,
e.g., the encoder and decoder for sparse code multiple access (SCMA), were
developed using a DNN based autoencoder in \cite{kim2018}.

One interesting characteristic of deep learning is that the DNN 
can be considered as a universal approximator \cite{Sun2017} which is capable 
of approximating an arbitrary function such that it can emulate the behavior of 
highly nonlinear and complicated systems. Moreover, given that the
DNN can be trained in an end-to-end manner which treats the
operation as a black box, i.e., end-to-end deep learning, the use of DNN enables the 
exploit of the optimal strategy without solving the complicated problems
explicitly, cf. Fig. \ref{fig_e-to-e-learn}. In this sense, the deep learning has been applied for the 
resource allocation (RA) of WCS where the exploitation of 
optimization problems was taken into account previously.
Unlike conventional approaches which derive the optimal strategy
for RA from the analytic system model with assumptions, the deep learning
based approach can derive the optimal strategy directly from actual channel
data such that it can adapt to the environment and 
the performance is likely to be higher in practice.
Moreover in the deep learning based approach, 
the general solver for the optimization problem of RA can be derived 
such that the optimal strategy can be found with low computation time 
even when the parameters, e.g., channel gain, change \cite{Ye2018, Lee2018}. 
The authors of \cite{Sun2017} used a DNN to 
regenerate the transmit power of weighted minimum mean 
square error (WMMSE)-based scheme in order to resolve 
the problem of high computation time of the WMMSE-based scheme.
Moreover, in \cite{Lee2018}, the transmit power was optimized to maximize
either spectral efficiency (SE) or energy efficiency (EE) 
where the optimal strategy for transmit power control is 
learned through an end-to-end deep learning without needing to 
derive explicit mathematical formulation.

In this article, we focus on the application of deep learning 
in the WCS. To this end, we first address basic principles of deep learning. 
Then, we investigate the DNN architecture which can be trained to
derive general strategy for RA that can achieve 
diverse goals, i.e., maximization of SE, EE, and minimization of 
transmit power, while satisfying constraints on interference and 
quality-of-service (QoS).
Finally, the research challenges regarding the application of 
deep learning will be addressed before concluding the article.

\section{Fundamentals of Deep Learning} 

In this section, we first describe the basic component and 
structure of Neural network (NN) and how to train it. 
Then, we turn our attention to 
deep learning and its application to WCS.
%==============================================================
\begin{figure}[t]
	\centerline{\includegraphics[width=7.0cm]{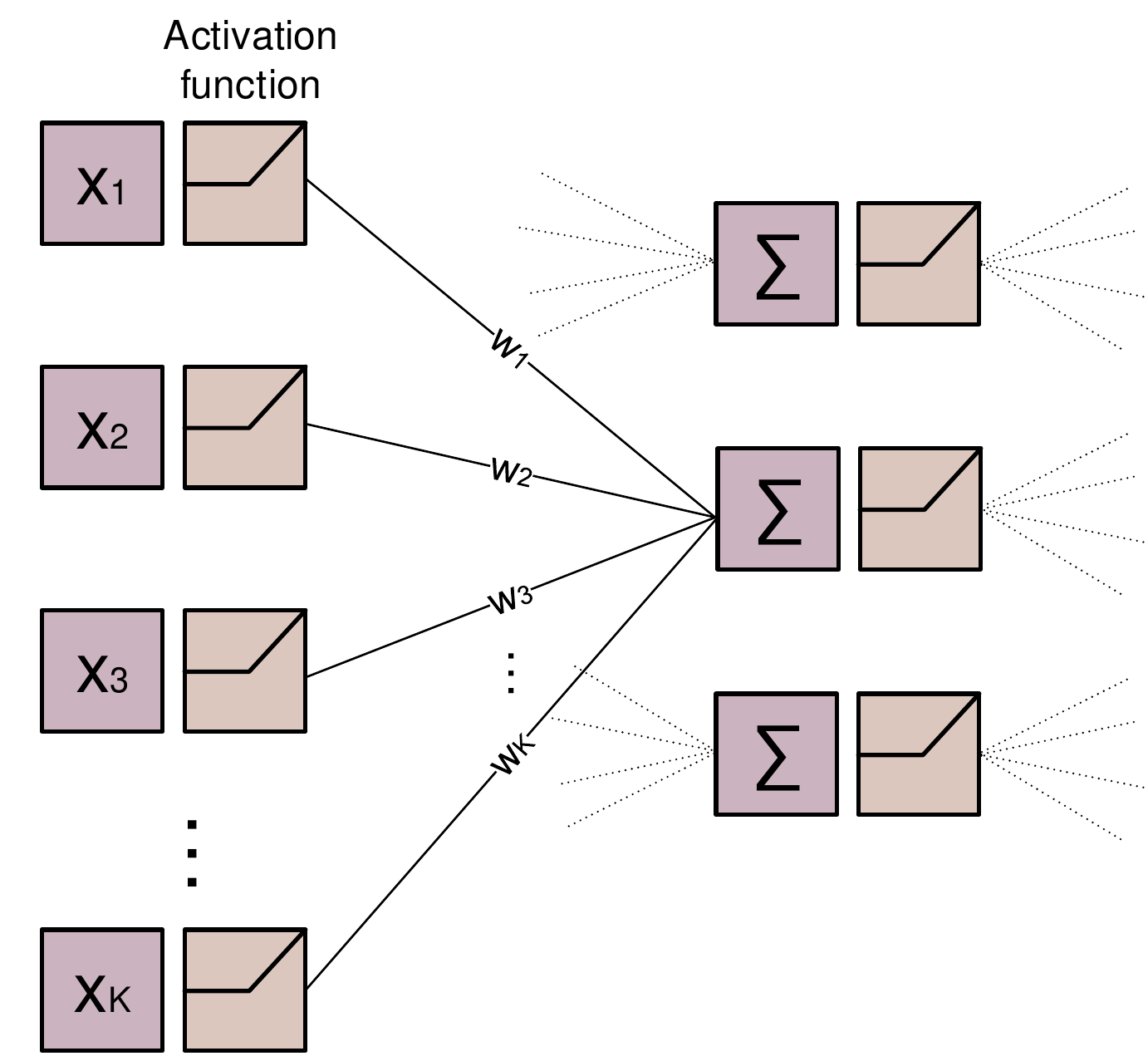}}
	\caption{Basic structure of NN.}
	\label{fig_nn_structure}
\end{figure}
%==============================================================

\subsection{Structure of Neural Network}

%% 아래 설명에서 ANN보다 NN이 좀 더 나을 것 같음
NN which comprises DNN, is a subcategory of 
machine learning which mimics the operation of brain. Through the extensive
experiment, it was found that the brain is composed of
neurons which are connected to each other. The neuron takes
the outputs of other neurons as its input and activates its 
output when the inputs satisfy a certain condition, i.e., it can be considered 
as a biological switch. In the NN, the connection 
between neurons (nodes) is modeled by the matrix multiplication and
the activation of neurons is modeled by the functions which 
look similar to a step function, e.g., sigmoid and rectified linear unit  (ReLU),
which provide the ability to model non-linear characteristic 
to the NN. The basic structure of NN is shown in Fig.\ref{fig_nn_structure}.
Accordingly, the NN can be considered as the 
collection of matrix computation and activation functions
and the calculation of output of NN for one sample data, 
i.e., inference, usually requires low computational overhead.

According to the existence of feedback loop in 
the network, the NN can be divided into two categories.

\begin{itemize}
\item \textbf{Feedforward neural network (FNN):} In FNN, there is no feedback loop 
such that the previous input data would not affect the current output, i.e., 
memoryless. The FNN is used when the data is not
correlated to each other, e.g., image data. In WCS, this type of 
NN is used for the spectrum sensing
and power control of users \cite{Sun2017, Lee2018, Lee2017}.

\item \textbf{Recurrent neural network (RNN):} The connection 
of nodes in RNN forms a cycle such that the current input
can affect the output of the next input, i.e., the DNN has memory. 
Accordingly, this type of NN is used for the data which 
has a temporal correlation. The data with 
temporal correlation in WCS, e.g., channel estimation, 
has been dealt with the RNN \cite{OShea2016a}.
\end{itemize}

\subsection{Overview on Training of Neural Network}

The training of NN, i.e., determination of weights
and biases of NN, is not obvious. In fact, the lack of
appropriate way to find weights and biases has brought the
first depression of research on NN after its first development in 1950s.
In 1980s, an efficient way to train NN, i.e., back-propagation algorithm, has 
been developed \cite{Rumelhart1987} which leads the second booming of the research on NN.
The back-propagation algorithm is based on the gradient descent 
technique such that the error of the NN, e.g., difference between
the target and the output, is propagated in backward direction
to update weights and biases according to the gradient, 
e.g., strengthen the parameters when the
output is correct and weakening the parameters when the output
is incorrect. By using the back-propagation algorithm, the NN 
can be trained efficiently.

According to the training methodology, the 
NN can be divided into three categories, same as general machine learning
algorithms.

\begin{itemize}
\item \textbf{Supervised learning:} In the supervised learning, 
the training data is labeled such that the NN can compare its
output with the ground truth. This type of learning is widely
used in the classification, e.g., the detection of primary user in cognitive 
radio (CR) systems \cite{Lee2017}.

\item \textbf{Unsupervised learning:} In the unsupervised learning, 
the data is not labeled such that the NN should autonomously 
derive the meaningful features from the input samples, e.g., 
clustering. In \cite{Kim2018b}, the encoder and decoder for SCMA system has been 
found using the autoencoder structure in unsupervised learning.

\item \textbf{Reinforcement learning:} In the reinforcement 
learning, the learning of NN is conducted by trial-and-error.
Especially, for a given input data, the proper action can be
found through NN and reward can be observed for the 
selected action. Then, the NN is trained to provide 
better action which gives higher reward. Anti-jamming strategy for secondary 
users in CR systems was developed based on deep 
reinforcement learning in \cite{Han2017}.
\end{itemize}

\subsection{Deep Neural Network}

With the back-propagation algorithm, the NN can be trained efficiently.
Nonetheless, it was observed that the NN with a large 
number of layers which is essential to achieve human-like
functionality, is hard to be trained, mainly due 
to the vanishing gradient problem, and the lack of proper input 
data and computation power, and it results in 
the second depression of research on NN for about 20 years.

Recently, the use of NN with a large number
of layers, which is known as DNN, becomes possible, mainly
due to the following four reasons \cite{LeCun2015}.

\begin{itemize}
\item \textbf{Availability of large dataset:} Due to the development
of sensors and Internet, the collection of a large number of data
is enabled which is essential for the DNN to learn general features.
For instance, the development of ImageNet, which is 
a database of image that contains more than 15,000,000 
labeled images, becomes the basis for the big success of DNN
in the image classification.

\item \textbf{Use of better activation function:} The problem of vanishing 
gradient is mainly caused by the use of inefficient activation functions whose
gradient is smaller than 1, e.g., sigmoid. However, 
in the DNN, the activation function with better gradient characteristics,
e.g., ReLU, is used such that the error at the output can be properly
propagated through the layers.
 
\item \textbf{Higher computation power:} Although the inference
of output for one input data requires small computational overhead, the training 
can take a long computation time due to the large number of training 
data. However, thanks to the development
of parallel computation based on graphics processing unit, this training procedure can be
conducted with low computation time.

\item \textbf{Efficient initialization methodology:} In the DNN, the initialization
technique is utmost important to achieve high performance without falling into
poor local minimum. Recently, good initialization techniques such as Restricted 
Boltzmann Machine (RBM) and Xavier initialization enable the efficient training of DNN 
\cite{Glorot2010}.

\end{itemize}

\subsection{Application of Deep Neural Network in Wireless Communications Systems}

The DNN is well suited for WCS because the WSC usually deals with a large amount of data 
and has complex system model which is hard to analyze at hand, and its potential application 
in WCS is enormous. Especially, unlike conventional approach in WCS where
the specific channel model is assumed such that 
its performance cannot be guaranteed when the assumed channel
is different from the actual channel, the DNN based approach is able to adapt 
its operation according to the environment without relying on the specific 
channel model. Moreover, same as image classification,
the DNN based approach might provide better schemes than conventional 
handcrafted schemes, especially, when the problem is complex and hard to analyze 
by empirical formulation \cite{kim2018, Lee2017, Lee2018}.

%%%%%% 딥러닝에 대한 기초적인 지식 전달 (그림 첨부?)
%%%%%% -> 슈퍼바이즈드 언슈퍼바이즈드??
%%%%%% -> 딥러닝 구조에 대한 설명?
%%%%%% -> 트레이닝 방안 설
%%%%%% -> 왜 딥러닝이 최근들어서 급상승인가? (ReLU, GPU, 데이터 량, initialization 등등)

%%% Intro에 설명함 %%%%%%%%%%%
%%%%%%%%%%%%%%%%%%%%%%%%%%%%%%%%%%%%%%%%%%%%%%%%%%%%%%%%%%%%%%%%%
%%%%%%%%%%%%%%%%%%%%%%%%%%%%%%%%%%%%%%%%%%%%%%%%%%%%%%%%%%%%%%%%%
%%%%%%%5\section{Application of Deep Learning in Communication Systems} 
%%%%%% 딥러닝이 왜 통신시스템에 적용되어야 하는가? (장점)
%%%%%% -> 빠른 연산, 블랙박스(general solver)를 만들기 때문에 입력이 바뀔때 새로 솔루션을 찾을 필요 없음
%%%%%% -> hand-craft 한 수식도출 필요 없음  (SCMA 인코더 디코더 언급), end-to-end learning
%%%%%% -> 또 다른 이유??
%%%%%%%%%%%%%%%%%%%%%%%%%%%%%%%%%%%%%%%%%%%%%%%%%%%%%%%%%%%%%%%%%
%%%%%%%%%%%%%%%%%%%%%%%%%%%%%%%%%%%%%%%%%%%%%%%%%%%%%%%%%%%%%%%%%

\section{Resource Allocation based on End-to-End Deep Learning} 

%%%%%% 우리 시뮬레이션 결과 간략히 보임 (CDF는 빼고 TX power, EE, rate, outage만 보임)
%%%%%% 2페이지 이내로 간략하게 작성

In this section, we investigate the DNN based RA.
To this end, we first describe the considered 
system model which comprises underlay device-to-device (D2D) 
communication. Then, the proposed DNN model for RA whose objective
is either maximizing SE, EE or minimizing total transmit power, is 
addressed and the optimality of the DNN based RA is examined
through simulation.

\subsection{System Model and Resource Allocation}

We consider an RA for the multi-channel cellular system 
with underlay D2D communication where the data transmission 
of $N$ D2D user equipments (DUEs) takes place simultaneously with 
cellular transmission, and users are randomly distributed over an 
area $D \times D$. The channel gain between $i$-th 
transmitter and $j$-th receiver for channel $m$ is denoted 
as $h^{m}_{i, j}$, where the index $0$ is assigned to the cellular user
equipment (CUE) and base station (BS).
The considered system model is depicted in Fig. \ref{fig_dnn_model}.
% W는 neuron의 w와 겹치기도하고 W와 N0는 수식에는 등장하지 않고 performance evaluation에만 
% 등장하는데 notation없이 그냥 performance evaluation에 로 글로 쓰기만하면 되지 않을까 싶음
%% => remove ok

% 그림이 남으면 deplyment에 대한 그림도 넣으면 좋을거 같음...
In the considered RA, the transmit power of DUEs allocated to each channel, 
which we denote as $P^{m}_{i}$ where $m$ and $i$ are the index of channel
and DUE, respectively, has to be determined to
either 1) maximize SE, 2) maximize EE, or 3) minimize total transmit power.
In the RA, we take into account
three constraints. First, the transmit power allocated to each channel 
should be non-negative and the sum of transmit power for a single 
DUE should not exceed the maximum transmit power, $P_{\textrm{T}}$
(i.e., transmit power constraint). Second, the amount of interference
caused to cellular transmission must be less than the 
threshold $I_{\textrm{T}}$ (i.e., interference constraint). 
Third, each D2D transmission should satisfy the minimum QoS
requirement, i.e., the SE achieved by each DUE has to be
larger than the threshold, $R_{\textrm{T}}$ (i.e., QoS constraint).

Given that the considered RA can be formulated
into a non-convex problem, it is hard to derive the 
optimal RA analytically, therefore, iterative methods based on 
Lagrangian relaxation have been considered previsouly \cite{Jiang2016}.
However, this approach requires a number of iterative computations, 
which possibly increases the computation time 
\cite{Sun2017, Lee2018}, 
such that the real time operation can be hindered. However, in the DNN based 
RA, the generic solver is derived autonomously through
DNN which involves only simple matrix operations, so that the 
proper RA for any channel condition can
be found with a low computational complexity without iteration.

\subsection{Resource Allocation based on Deep Neural Network}

\begin{figure*}[t]
	\centerline{\includegraphics[width=16.0cm]{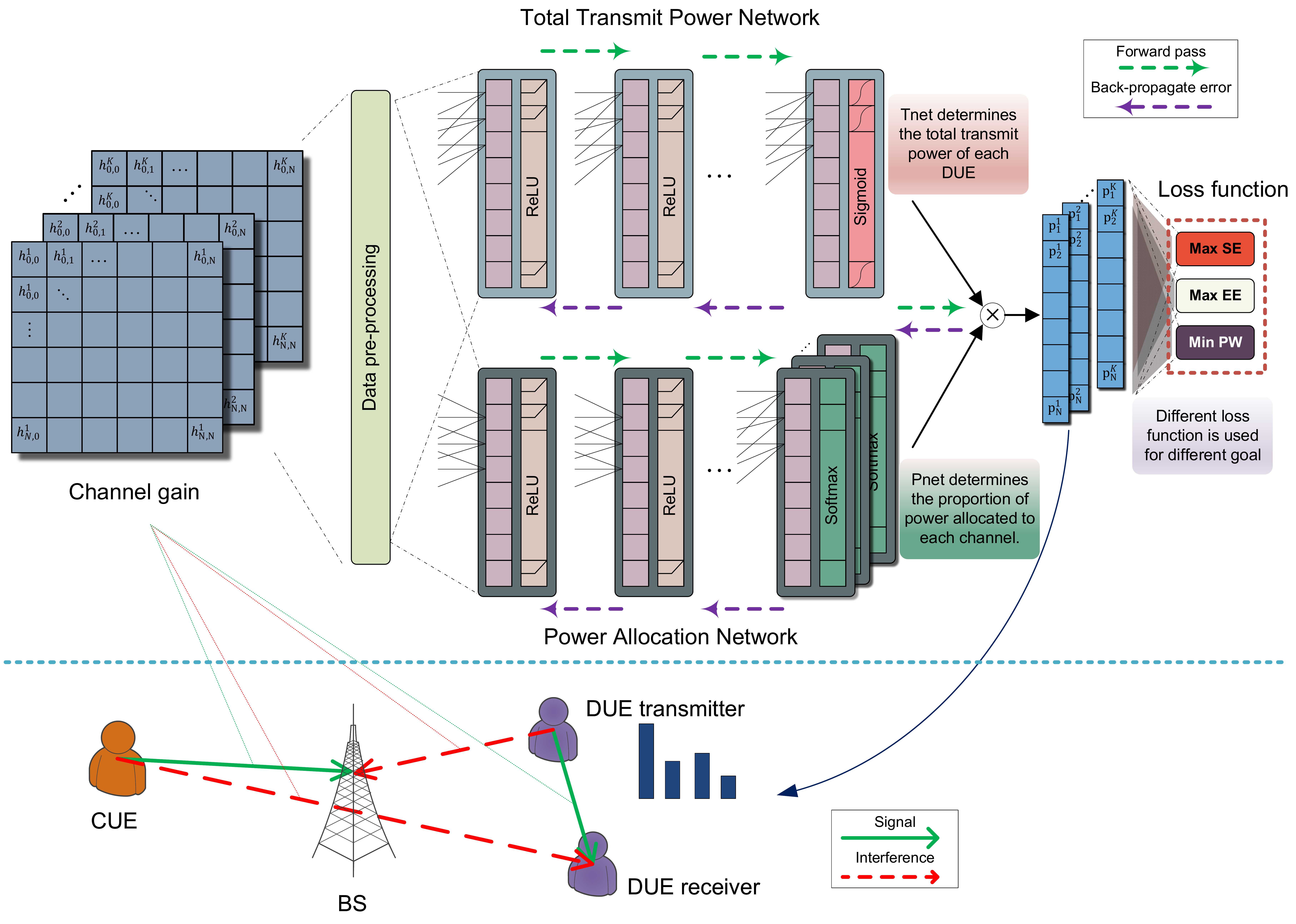}}
	\caption{Considered system model and DNN structure for the RA.}
	\label{fig_dnn_model}
\end{figure*}

In our DNN model, the total transmit power of each DUE and 
the proportion of transmit power allocated to each channel
by individual DUE are found jointly using separate DNN 
module, which are the total transmit power network (Tnet)
and the power allocation network (Pnet), cf. Fig. \ref{fig_dnn_model}. 
The channel gain is pre-processed for better training performance
such that it is converted to dB and normalized to have zero mean and unit variance, and these
pre-processed channel gain becomes the input of two networks, Tnet and Pnet.
%==============================================================
%==============================================================

The Tnet and Pnet are composed of sub-modules which comprised
a FC layer and ReLU which is used as an activation 
function. Given that the each output of the Tnet should be less than the 
maximum transmit power, $P_{\textrm{T}}$, we have implemented
the sigmoid which is multiplied by the $P_{\textrm{T}}$ at the end of the Tnet, where the output
of sigmoid is between 0 and 1. On the other hand, the proportion of transmit power allocated
to each channel for each DUE is determined by the Pnet, 
such that we consider $N$ softmax modules as the last 
layer of Pnet where each softmax module has $M$ outputs.
Given that the sum of a single softmax module is 1, it is appropriate 
to model the proportion of transmit power allocated to channels
for individual DUE. Finally, by multiplying the outputs of
both Tnet and Pnet, the transmit power of DUEs allocated to each channel 
can be found. It should be noted that the same DNN structure 
with different values of weights and biases can be used for three
different objectives such that it is efficient
in practice in terms of reusing the same DNN structure.

In order to find the optimal set of weights and biases, our proposed DNN 
model has to be trained first. To this end, the channel samples 
for training must first be collected by either measurement or simulations using 
well-known channel model. Hybrid use of measurement data and synthetic data, 
where the DNN is initially trained with synthetic channel data, and then, trained 
with a few actually measured data for fine tuning, which is known as 
a transfer learning, is also possible.

% 앞에서 weighted sum이라고 언급해서 람다들은 weight라고 표현하는 것이 
%낫지 않을까요? 그리고 어차피 개념적인걸 설명할거면 델타를 빼도 좋지 않을까요?
%%% => 정확히 weight는 아니므로 다른식으로 명명하는 것이 좋을 것 같음
After the channel samples are prepared, the proposed DNN can be trained through a back-propagation 
algorithm. For the training, in order to achieve goal 
while satisfying constraints, the weighted sum of objective function
and the functions on constraints is taken into account. 
Accordingly, in order to train DNN model to maximize 
the SE, the loss function, $L_{\textrm{SE}}$, can be set as 
$L_{\textrm{SE}} = -   \sum \textrm{SE}_{i} +  \lambda_1  \sum \tanh \left([ I^{k}_{\textrm{CUE}}   - I_{\textrm{T}}]^{+} \right) +  \lambda_2  \sum \tanh \left([ \textrm{SE}_{i}  - R_{\textrm{T}}]^{+} \right)$, 
where $ \lambda_1$,  $ \lambda_2$ are the controlling parameters, 
$\textrm{SE}_{i}$ is the SE of DUE $i$, and $I^{k}_{\textrm{CUE}}$ is the interference
caused to CUE at channel $k$. 

As can be seen from the formulation, the loss function 
increases as the SE of DUEs decreases and at the same time when the interference and 
QoS constraints\footnote{Given that the transmit power constraint is 
always satisfied in our DNN model, we do not consider this constraint in the loss 
function.} are not satisfied. Given that the DNN is trained to reduce the loss, 
through training, the SE of DUEs will be increased and the violation of interference
and QoS constraint will be reduced. Note that $\lambda_1$ and
$\lambda_2$ determine\footnote{For example, when 
the value of $\lambda_1$ is small, the interference constraint
is barely considered in the loss function such that 
the transmit power is learned without
considering the inference caused to the CUE.} 
the penalty for the violation of constraints.

Similarly, the loss functions to maximize EE, which we denote as 
$L_{\textrm{EE}}$ can be written
as $L_{\textrm{EE}} = -   \sum \textrm{EE}_{i} +  \lambda_1  \sum \tanh \left([ I^{k}_{\textrm{CUE}}   - I_{\textrm{T}}]^{+} \right) +  \lambda_2  \sum \tanh \left([ \textrm{SE}_{i}  - R_{\textrm{T}}]^{+} \right)$
and the loss function to minimize the total transmit power, which we denote as $L_{\textrm{TP}}$, 
can be written as $L_{\textrm{TP}} =   \sum P^{m}_{i} +  \lambda_1  \sum \tanh \left([ I^{k}_{\textrm{CUE}}   - I_{\textrm{T}}]^{+} \right) +  \lambda_2  \sum \tanh \left([ \textrm{SE}_{i}  - R_{\textrm{T}}]^{+} \right)$.
Although the considered loss functions
are different from loss functions which are commonly
considered in the deep learning researches, e.g., cross entropy, 
the back-propagation based learning is still possible because
all the loss functions are differentiable.

After the proper loss function is determined, 
the training of DNN can be conducted efficiently using 
off-the-shelf stochastic gradient descent algorithms,
e.g., Adaptive Moment Estimation algorithm. 
After training, the appropriate transmit power allocated to each 
channel can be determined by feeding the current channel
gain to the trained DNN.

\subsection{Performance Evaluation}

We compare the performance of the proposed DNN based RA with the
optimal performance which is found through 
an exhaustive search. For the performance evaluation, 
we assume that the number of DUE transmission pairs 
and the number of channels is two such that the finding optimal performance 
through exhaustive search is computationally plausible. 
The maximum transmit power and the circuit power 
are set to 23 dBm, the bandwidth is set to 10 MHz, 
the noise density is set to -173dBm/Hz, $I_{\textrm{T}}$ = -55 dBm and  
$R_{\textrm{T}}$ = 3 bps/Hz. Moreover, the simplified path loss model 
with path loss coefficient $10^{3.453}$ and
path loss exponent $3.8$ is considered and an 
independent and identically distributed (i.i.d.) circularly symmetric 
complex Gaussian (CSCG) random variable with a mean of zero 
and a variance of one is used for multipath fading.

For the DNN model, we assume that the number of layers for 
Tnet and Pnet is 4 and the number of hidden nodes for FC layer
is 100. Moreover, 40,000 channel samples are used for the training 
and the learning rate are 
adaptively changed over training epoch.
In the performance evaluation, we examine the average SE, 
the average EE, and the total transmit power of each DUE of
the DNN based RA whose objectives are the maximization of 
SE (Max. SE), the maximization of EE (Max. EE) and the 
minimization of total transmit power (Min. PW). 
Moreover, the optimal performance of RA for 
considered goals is also examined.
Although we do not show graphically, the computation time of 
the DNN based RA was measured at 1.1 milliseconds while 
the time required to find the optimal solution is measured 
at 2743 milliseconds which reveals the benefit of the 
DNN based RA.
%==============================================================
\begin{figure}
\centerline{\includegraphics[width=7.0cm]{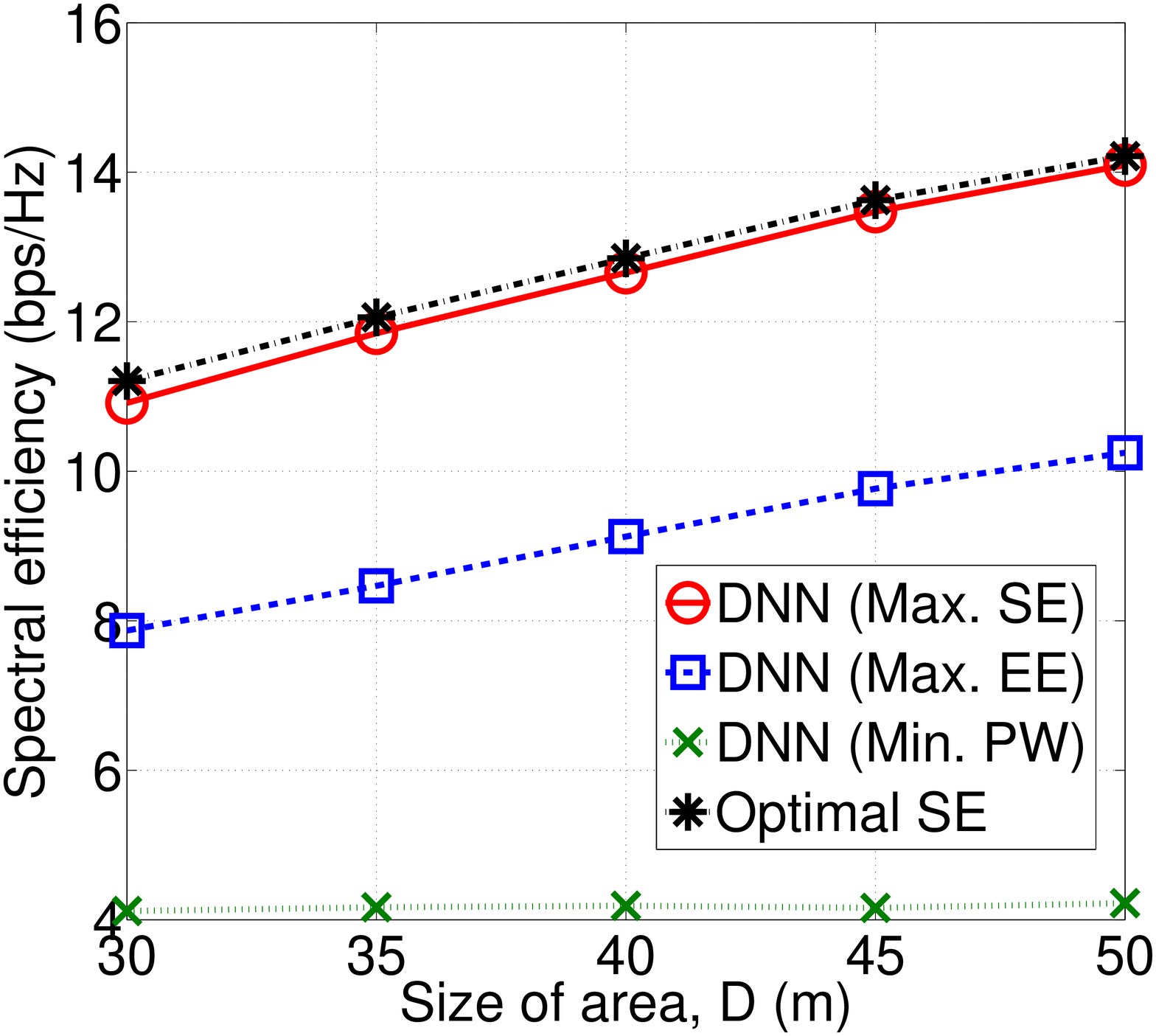}}
 \caption{Spectral efficiency vs. size of area.}
\label{fig_rate_var_size}
\end{figure}
%==============================================================
%==============================================================
\begin{figure}
\centerline{\includegraphics[width=7.0cm]{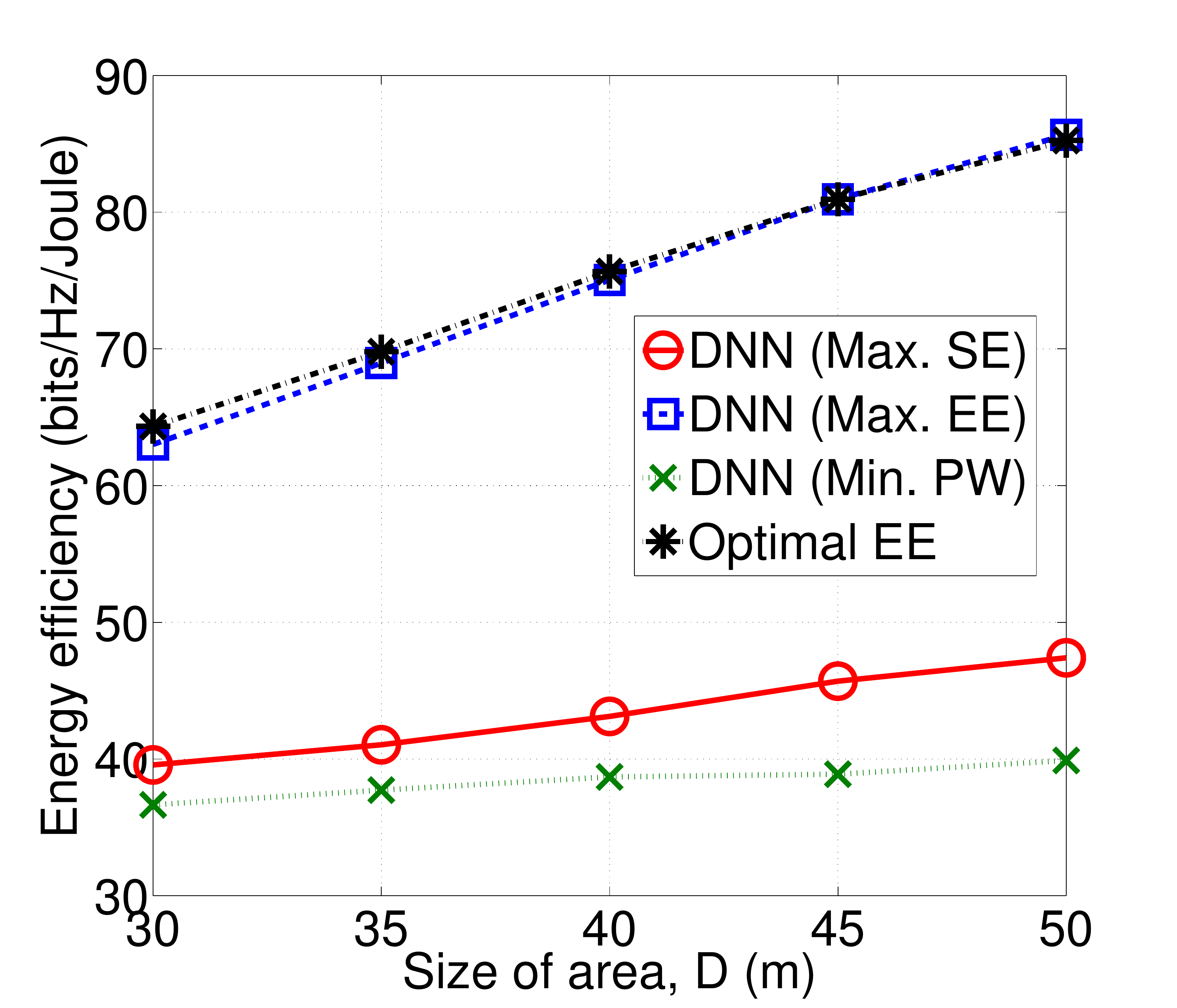}}
 \caption{Energy efficiency vs. size of area.}
\label{fig_ee_var_size}
\end{figure}
%==============================================================
%==============================================================
\begin{figure}
\centerline{\includegraphics[width=7.0cm]{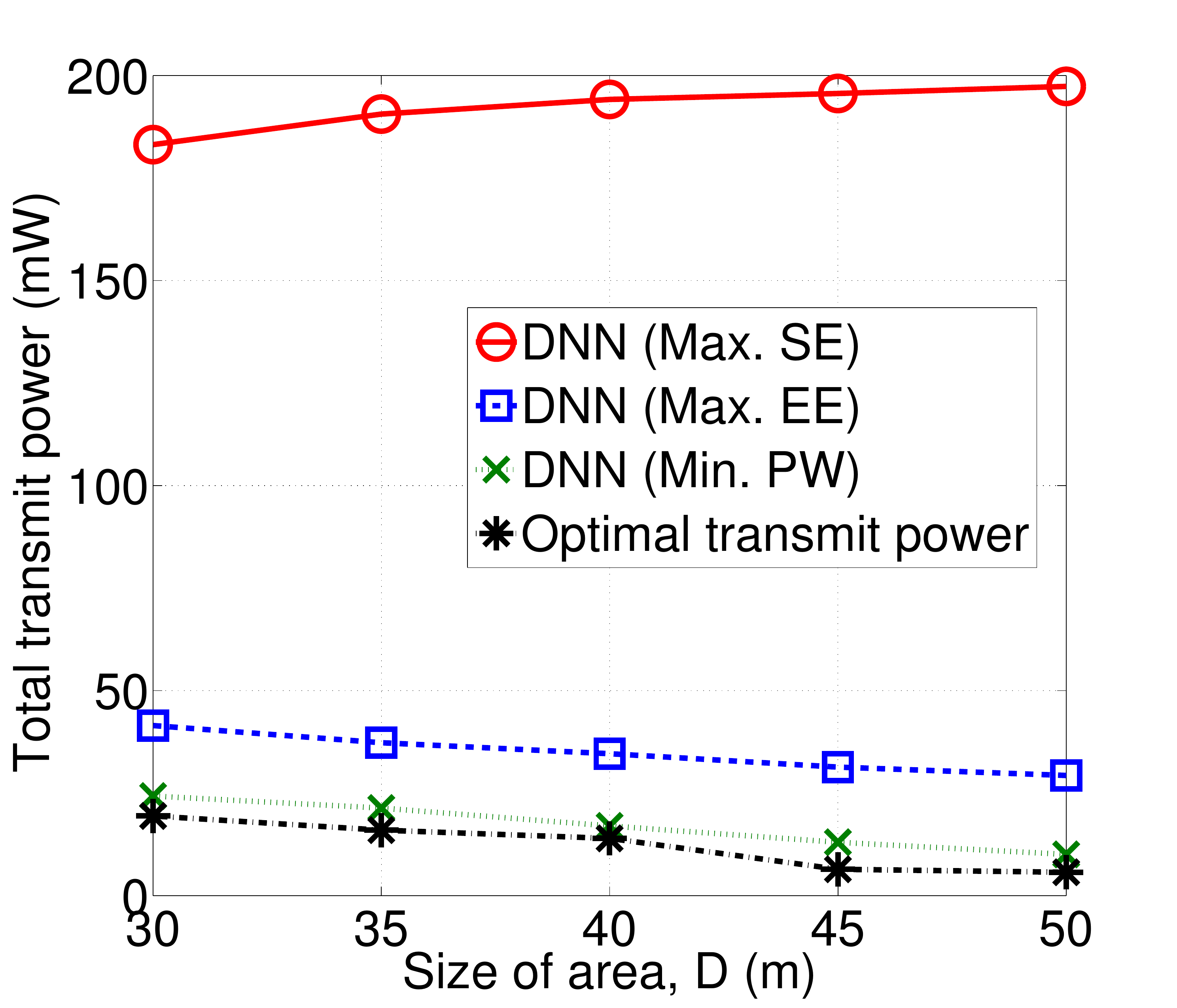}}
 \caption{Total transmit power vs. size of area.}
\label{fig_pw_var_size}
\end{figure}
%==============================================================

In Figs. \ref{fig_rate_var_size}, \ref{fig_ee_var_size}, and \ref{fig_pw_var_size}, 
we show the average SE, the average EE, and the total transmit
power of DUE, respectively, as a function of the size of area, $D$.
As can be seen from the simulation results, the DNN based RA achieves the 
close-to-optimal performance. For example, the SE of DNN based RA 
for the SE maximization achieves 97\% of the optimal SE. 
The outage probability that interference or 
QoS constraint is violated is measured to be less than 2.1\%
which is sufficiently low. Moreover, even when the constraint
is violated, the level of violation is minor. These simulation results affirm 
the applicability and benefits of the DNN based RA because the 
DNN based RA has much lower computation time
compared with the optimal scheme. 

\section{Research Challenges} 

In the following, we discuss some research challenges 
for future usage of deep learning in WCS.

\subsection{Measured Channel Data as Training Sample}

In the deep learning based approach, the DNN 
autonomously finds the optimal strategy or meaningful features
directly from data instead of using handcrafted
mathematical system model. Accordingly,
in order to achieve high performance in deep learning, 
a large number of training samples regarding WCS, e.g., 
channel gain data, for different scenarios has to be collected
through actual measurement\footnote{For example, the large set of 
labeled image data, i.e., ImageNet dataset, enables the big success of deep learning
technology in the image classification.}.
The problem regarding the preparation of 
training data could be solved by using the DNN 
based on a generative model, e.g., generative adversarial 
network (GAN), which shows big success in the image 
synthesis, to generate realistic synthetic training data.

%%%%%%%%%
%%%
%% 내용추가 - 딥러닝을 쓰면 채널 모델없이 정말 어댑티브하게 성능개발 가능

\subsection{Distributed Operation and Inaccurate Channel Information}

Unlike image classification 
in which all input data can be easily obtained by the single node, in 
WCS, the input data of the DNN, e.g., channel gain, is hard to 
be obtained by the single node which executes the DNN functionality
due to high signaling overhead, especially when the 
number of users is large. Moreover, the input data of 
DNN for practical WCS is likely to contain error due to the inaccurate
measurement and delayed channel feedback. To solve the
problem of distributed operation, the distributed deep learning
can be considered \cite{Kerret2018} and to solve the problem 
of inaccurate channel information, denoising autoencoder 
can be taken into account.

\subsection{Computation Complexity}

The control of WCS, e.g., RA, has to be conducted within 
a very short time period, i.e., several milliseconds, due to the short
frame length. Moreover, the deep learning based scheme for
WCS should have low computational complexity because
it could be operated on the mobile device which has limited
computation power. Although we show that the DNN based
RA has a very low computation time compared with the optimal scheme,  
storing the trained DNN model can become overhead
to the mobile device. Moreover, online learning in which each 
mobile device trains its own model based on the collected
channel samples, can be taken into account in order to further 
improve the performance, and in this case, the overhead
of DNN based schemes can be large. This problem can be
solved by using newly developed AI accelerator chip,
e.g., tensor processing unit (TPU) by Google, which is
likely to be implemented in mobile device because more 
technologies based on deep learning are now
applied to mobile devices.

\section{Conclusions}

In this article, the application of DNN for WCS 
through autonomous end-to-end deep learning as opposed to 
conventional handcrafted engineering based on  
the mathematical modeling, has been discussed. In particular, 
the deep learning based RA has been addressed whose optimal 
strategy is hard to be obtained in conventional approach. 
It has been confirmed by performance evaluation that the DNN based RA 
can achieve close-to-optimal performance with low computation
time for various objectives of RA reusing the same DNN structure. 
We have also outlined the challenges regarding the application of deep learning
in the research of WCS.

\bibliographystyle{IEEEtran}
\bibliography{IEEEabrv,mybibfile}

\begin{thebibliography}{10}
\providecommand{\url}[1]{#1}
\csname url@samestyle\endcsname
\providecommand{\newblock}{\relax}
\providecommand{\bibinfo}[2]{#2}
\providecommand{\BIBentrySTDinterwordspacing}{\spaceskip=0pt\relax}
\providecommand{\BIBentryALTinterwordstretchfactor}{4}
\providecommand{\BIBentryALTinterwordspacing}{\spaceskip=\fontdimen2\font plus
\BIBentryALTinterwordstretchfactor\fontdimen3\font minus
  \fontdimen4\font\relax}
\providecommand{\BIBforeignlanguage}[2]{{%
\expandafter\ifx\csname l@#1\endcsname\relax
\typeout{** WARNING: IEEEtran.bst: No hyphenation pattern has been}%
\typeout{** loaded for the language `#1'. Using the pattern for}%
\typeout{** the default language instead.}%
\else
\language=\csname l@#1\endcsname
\fi
#2}}
\providecommand{\BIBdecl}{\relax}
\BIBdecl

\bibitem{Shea2016}
T.~J. {O'Shea}, J.~Corgan, and T.~C. Clancy, ``Convolutional radio modulation
  recognition networks,'' in \emph{Proc. of EANN}, Aberdeen, U.K., Sep. 2016.

\bibitem{Karpathy2015}
A.~Karpathy and L.~Fei-Fei, ``Deep visual-semantic alignments for generating
  image descriptions,'' in \emph{Proc. of IEEE CVPR}, Boston, MA, USA, Jun.
  2015.

\bibitem{LeCun2015}
Y.~LeCun, Y.~Bengio, and G.~Hinton, ``Deep learning,'' \emph{Nature}, vol. 521,
  no. 7553, pp. 436--444, May 2015.

\bibitem{OShea2016a}
T.~J. O'Shea, S.~Hitefield, and J.~Corgan, ``End-to-end radio traffic sequence
  recognition with deep recurrent neural networks,'' \emph{arXiv preprint
  arXiv:1610.00564}, 2016.

\bibitem{Ye2018}
H.~Ye, G.~Y. Li, and B.~H. Juang, ``Power of deep learning for channel
  estimation and signal detection in {OFDM} systems,'' \emph{{IEEE} Wireless
  Commun. Lett.}, vol.~7, no.~1, pp. 114--117, Feb. 2018.

\bibitem{kim2018}
M.~Kim, N.~I. Kim, W.~Lee, and D.~H. Cho, ``Deep learning aided {SCMA},''
  \emph{{IEEE} Commun. Lett.}, vol.~22, no.~4, pp. 720--723, Apr. 2018.

\bibitem{Sun2017}
H.~Sun, X.~Chen, Q.~Shi, M.~Hong, X.~Fu, and N.~D. Sidiropoulos, ``Learning to
  optimize: Training deep neural networks for wireless resource management,''
  \emph{arXiv preprint arXiv:1705.09412}, 2017.

\bibitem{Lee2018}
W.~Lee, M.~Kim, and D.~H. Cho, ``Deep power control: {T}ransmit power control
  scheme based on convolutional neural network,'' \emph{{IEEE} Commun. Lett.},
  vol.~22, no.~6, pp. 1276--1279, 2018.

\bibitem{Lee2017}
W.~Lee, M.~Kim, and D.-H. Cho, ``Deep sensing: {Cooperative} spectrum sensing
  based on convolutional neural networks,'' \emph{arXiv preprint
  arXiv:1705.08164}, 2017.

\bibitem{Rumelhart1987}
D.~E. Rumelhart and J.~L. McClelland, \emph{Learning Internal Representations
  by Error Propagation}.\hskip 1em plus 0.5em minus 0.4em\relax MIT Press,
  1987, pp. 318--362.

\bibitem{Kim2018b}
M.~Kim, N.~I. Kim, W.~Lee, and D.~H. Cho, ``Deep learning aided {SCMA},''
  \emph{{IEEE} Commun. Lett.}, vol.~22, no.~4, pp. 720--723, Apr. 2018.

\bibitem{Han2017}
G.~Han, L.~Xiao, and H.~V. Poor, ``Two-dimensional anti-jamming communication
  based on deep reinforcement learning,'' in \emph{Proc. of IEEE ICASSP}, New
  Orleans, LA, USA, Mar. 2017.

\bibitem{Glorot2010}
X.~Glorot and Y.~Bengio, ``Understanding the difficulty of training deep
  feedforward neural networks,'' in \emph{Proc. of AISTATS}, Sardinia, Italy,
  May 2010.

\bibitem{Jiang2016}
Y.~Jiang, Q.~Liu, F.~Zheng, X.~Gao, and X.~You, ``Energy-efficient joint
  resource allocation and power control for {D2D} communications,''
  \emph{{IEEE} Trans. Veh. Technol.}, vol.~65, no.~8, pp. 6119--6127, Aug.
  2016.

\bibitem{Kerret2018}
P.~de~{K}erret and D.~{G}esbert, ``{R}obust decentralized joint precoding using
  team deep neural network,'' in \emph{Proc. of ISWCS}, {L}isbon, Portugal,
  Aug. 2018.

\end{thebibliography}

%%%%%% 레퍼런스를 15개 이내 (보통 15개 채우는 것 같음)

\end{document}